\documentclass[pre,showpacs,twocolumn,superscriptaddress,floatfix]{revtex4}
\usepackage{amssymb,amsfonts,amsmath}
\usepackage[T1]{fontenc}
\usepackage{lmodern}
\usepackage{graphicx}
\begin{document}
%\graphicspath{{../figures/Abbildungen_neu/}}
\title{Cluster virial expansion for the equation of state of partially ionized hydrogen plasma}
\author{ Y. A. Omarbakiyeva}
\email{yultuz@physics.kz}
\affiliation{Institute of Physics, University of Rostock, D-18051, Rostock, Germany}
\affiliation{IETP, Al-Farabi Kazakh National University, 96a, Tole bi St., Almaty 050012, Kazakhstan}
\author{C. Fortmann}
\affiliation{Department of Physics and Astronomy, University of
California Los Angeles, Los Angeles CA 90095, USA}
\author{T. S. Ramazanov}
\affiliation{IETP, Al-Farabi Kazakh National University, 96a, Tole bi St., Almaty 050012, Kazakhstan}
\author{G. R\"{o}pke}
\affiliation{Institute of Physics, University of Rostock, D-18051, Rostock, Germany}
\date{\today}

\begin{abstract}
 We study the contribution of electron-atom interaction to the equation of state for partially ionized
 hydrogen plasma using the cluster-virial expansion. For the first time, we use the Beth-Uhlenbeck approach to calculate
 the second virial coefficient for the
 electron-atom (bound cluster) pair from the corresponding scattering phase-shifts and binding energies. 
 Experimental scattering cross-sections as well as phase-shifts calculated on the basis of different pseudopotential models are used
 as an input for the Beth-Uhlenbeck formula. By including Pauli blocking and screening in the phase-shift calculation, 
 we generalize the cluster-virial expansion in order to cover also near solid density plasmas.
 We present results for the electron-atom contribution to the
 virial expansion and the corresponding equation of state, i.e. pressure, composition,
 and chemical potential as a function of density and temperature.
 These results are compared with semi-empirical approaches to the thermodynamics of partially ionized plasmas.
Avoiding any ill-founded input quantities, the Beth-Uhlenbeck second virial coefficient for the electron-atom
interaction represents a benchmark for other, semi-empirical approaches.
\end{abstract}
\maketitle
%
%%%%%%%%%%%%%%%%%%%%%%%%%%%%%%%%%%%%%%%%%%%%%%%%%
\section{Introduction\label{sec:Introduction}}

The thermodynamics of dense plasmas, in particular hydrogen, has become an important topic. Models
of planetary interiors and stars depend on the equation of state (EOS) of the most abundant elements \cite{daeppen}. Reliable
EOS data for hydrogen are indispensable for the planning and conduction of
inertial confinement fusion experiments \cite{Lindl_PoP11_339_2004}. Within the quantum statistical approach to the EOS,
numerical techniques like density-functional theory and quantum molecular dynamics simulations have been elaborated
\cite{FehskeSchneiderWeisseCompMPP_2008}. Alternatively, analytical
approaches have been derived using many-body perturbation theory \cite{kremp} that yield, e.g., rigorous results for limiting cases, such as the
low density limit, that are hardly accessible by numerical methods.

A rigorous result for systems with short-range interaction is the Beth-Uhlenbeck formula that expresses the second virial coefficient
in terms of the scattering phase-shifts and the bound state energies \cite{BU}. Some generalizations have been performed to obtain results for a
larger range of densities by including medium effects, see \cite{kraeft,ZKKKR} for charged particle systems or \cite{roepke_bu} for nuclear systems.
Another particularly successful method to investigate strongly correlated many-particle systems is the
cluster-virial expansion \cite{horowitz}. In addition to the elementary constituents, the different bound states (cluster)
are considered as new reacting species in thermal equilibrium. The thermodynamic properties are expanded in terms of the
fugacities of the various components of the system and cluster-virial coefficients are introduced. In the case of nuclear 
systems \cite{horowitz,R,T}, one can consider the nucleons, deuterons (two-body cluster), and alpha-particles
(four-body cluster) as well as possible further nuclei as the components of the system. For ionic plasmas, electrons, ions, atoms, 
and molecules
are such components. Another example is the electron-hole exciton system \cite{KraeftKilimannKremp, Zimmermann}.

The cluster-virial expansion represents a ``chemical'' picture in the sense that the
virial is expanded in orders of the fugacities of the different components (single-particle states, bound states) in the system. 
In contrast, the traditional virial expansion is a ``physical'' picture, i.e.
the fugacities of elementary particles, such as electrons and nuclei, are the expansion variables. 
In the physical picture, bound states appear in higher order virial coefficients; their treatment involves sophisticated mathematics.
On the other hand, bound states are naturally included already in the lowest order of the chemical picture.
For special parameter values, bound state formation gives the leading contribution, e.g. in atomic or molecular gases.
The chemical picture accounts for these main terms already in the lowest order of the virial expansion,
whereas within the physical picture we have to consider higher orders of the expansion to identify the leading contributions.

We consider a partially ionized hydrogen plasma, which consists of three components as electrons ($e$), ions ($i$),
and hydrogen atoms ($a$).
The formation of heavier clusters, such as molecules or molecular
ions, can also be included but this is not considered in the present work.
Restricting ourselves to these three components, the relevant interactions are the elementary Coulomb interaction ($e-i,\,e-e,\,i-i$)
and the more complex interaction ($e-a,\,i-a,\,a-a$) with the atoms.
The Coulombic contributions to the EOS have been investigated extensively elsewhere, see Ref. \cite{kraeft}, and will not be given here. 

The treatment of electron-atom interaction in EOS studies is still an open question. 
A widely used method consists in constructing an effective electron-ion potential.
The pseudopotential method allows to include medium effects and thereby to 
enlarge the region of applicability of the cluster-virial expansion.
Such pseudopotentials should reproduce calculated phase-shifts and experimental scattering data.
The disadvantage of the pseudopotential method is that 
it depends on a local, energy-independent potential 
that can be introduced only in certain approximations to replace the energy dependent three-particle 
T-matrix. Well-known examples are the Buckingham potential \cite{rrz}, see also more recent approaches \cite{RDO}.
Other semiempirical methods, such as the excluded volume concept \cite{EbFortov} discussed below, 
depend on free parameters that lack a proper quantum-statistical foundation.

In this work, we overcome the shortcomings of the pseudopotential method by deriving
results for the electron-atom virial coefficient that are based on experimental data for the
electron-atom scattering cross-section. Therefore,
our results represent benchmarks for the electron-atom contribution to the EOS, only limited by the accuracy of measured scattering data.
We compare our results to different pseudopotential calculations.
As an example, we construct a separable
pseudopotential for the electron-atom interaction, and consider medium effects such as self-energy and Pauli blocking.

We present results for the EOS, in particular the pressure, composition, and chemical potentials for temperatures in the range of
$T\leq 10^{5}\,\text{K}$ and for densities in the order of $n\leq 10^{21}\,\text{cm}^{-3}$, where $n=n_\mathrm{e}^{\rm total}=n_\mathrm{i}^{\rm total}$
denotes the total number density of electrons or ions, respectively, that are found in free or bound states,
and obey the neutrality condition. For these parameter values, the hydrogen plasma is moderately coupled with the plasma parameter
$\Gamma= e^2/(4\pi\epsilon_0k_\mathrm{B}Td)<1$,
where $d=(3/4\pi n)^{1/3}$ is the average interparticle distance.
Effects of Fermi degeneracy are weak for $\Theta = k_\mathrm{B}T/E_\mathrm{F}\geq 1$, with the Fermi energy $E_\mathrm{F}=\hbar^2 (3\pi^2 n)^{2/3}/2m_\mathrm{e}$ where, more consistently, only the density of free electrons should be taken.
With increasing density and/or decreasing temperatures, the system becomes degenerate.

The present work is organized as follows: In Sec~\ref{sec:virial_expansion}, we briefly review the cluster-virial expansion
and the Beth-Uhlenbeck formula for the second virial coefficient. Section~\ref{sec:phaseshifts}
contains the calculation of the scattering
phase-shifts for the electron-atom system, both via experimental differential cross sections and phase-shifts from appropriate
pseudopotentials. In Sec.~\ref{sec:2ndvc}, the phase-shifts
are used to calculate the corresponding second virial coefficient.
Based on these results, corrections for the pressure and the chemical potentials are calculated in Sec.~\ref{sec:thermodynamics}.
Sec.~\ref{sec:PauliBlock} contains the presentation of the generalized Beth-Uhlenbeck
approach and the calculation of the density dependent second virial coefficient.
Conclusions are drawn in Sec.~\ref{sec:conclusion}.

%%%%%%%%%%%%%%%%%%%%%%%%%%%%%%%%%%%%%%%%%%%%%%%%%%%%%%%%%%%%%%%
\section{Cluster virial expansion and the Beth-Uhlenbeck formula\label{sec:virial_expansion}}
The canonical partition function of an interacting many-particle system at temperature $T$,
volume $V$, composed of $N_{c}$ particles per species $c$ ($c=e,\,i,\,a$) carrying spin $s_c$, reads
\begin{eqnarray}
\label{eq:1}
Z^{\rm can}(T,V,N_{c})=\mathrm{Tr}\,\{\exp(-\beta H)\}~,
\end{eqnarray}
with $\beta=1/k_\mathrm{B}T$. Here and in the following, the spin of particles of species $c$ is implicitely
taken as $s_e=1/2$ for electrons, $s_i=1/2$ for protons 
and for atoms in the singlet state (antiparallel spins of electron and proton) $s_{a, \rm singlet}=0$, and
$s_{a, \rm triplet}=1$ for the triplet state (parallel spins of $e$ and $i$).
We neglect hyperfine splitting of the atomic levels. 
   
The
Hamiltonian
\begin{eqnarray}
\label{eq:2}
H=
\sum_{c}\sum_{j=1}^{N_{c}}&&\left[E^{(0)}_c+ \frac{p_{j}^2}{2 m_{c}} \right]
\nonumber\\
&&+\frac{1}{2}\sum_{cd}\sum_{j=1}^{N_{c}}\sum_{k=1}^{N_{d}}{}'\,\, V_{cd}(\vec{r}_{j}-\vec{r}_{k})
\end{eqnarray}
contains, besides the kinetic energy of each particle, the mutual interaction,
represented by the two-particle interaction potential
$V_{cd}(\vec r_j-\vec r_k)$. The prime indicates that self-interaction is excluded,
and the energies for each component are gauged to $E^{(0)}_c$ if the particles are at rest.
From a relativistic approach, $E^{(0)}_c$ is given by the rest mass containing
the binding energy. In our nonrelativistic approach,
we choose the gauge relatively to the rest mass of the elementary particles
so that $E^{(0)}_c$ is the binding energy of the composites.

Using the Mayer cluster expansion \cite{huang},
we arrive at the virial expansion for the free energy $F=-k_\mathrm{B}T\,\ln Z^{\rm can}$ valid for short-range potentials,
\begin{eqnarray}
\label{eq:3}
F(T,V,N_{c})=F_{\rm id}(T,V,N_{c})-k_{\rm B}TV\Big\{ \sum_{cd}n_{c}n_{d}b_{cd}\nonumber \\
+\sum_{cde}n_{c}n_{d}n_{e}b_{cde}+...\Big\}\,.
\end{eqnarray}
$F_{\rm id}(T,V,N_{c})=k_{\rm B}T\sum_{c}N_{c}\left[\ln\left(\frac{\Lambda_{c}^3
N_{c}}{V g_c}\right)-1+\beta E^{(0)}_c\right]$ is the free energy for the classical ideal
(i.e. non-interacting) system, where $\Lambda_c=\left( 2\pi\hbar^2/k_\mathrm{B}Tm_c \right)^{1/2}$ denotes the
thermal wavelength of species $c$, and $n_c=N_c/V$ the particle number density of the component $c$,
$g_c=2s_c+1$ is the spin degeneracy factor. 
The expansion coefficients
$b_{cd}$ and $b_{cde}$ are the second and third virial coefficients, respectively.
They are determined by the interaction, but also by degeneracy terms.

Having the virial coefficients at our disposal,
we can easily derive the thermodynamic properties of the system under consideration. E.g. for the
pressure and the chemical potential the following expressions are found using the standard thermodynamic relations:
\begin{eqnarray}
\label{eq:4}
p(T,V,N_{c})=p_{\rm
id}(T,V,N_{c})-k_{\rm B}T\Big\{\sum_{cd}n_{c}n_{d}b_{cd}
\nonumber\\
+2 \sum_{cde}n_{c}n_{d}n_{e}b_{cde}+...\Big\}~,\\
\label{eq:5}
\mu_{c}(T,V,N_{c})=\mu_{c,\rm
id}(T,V,N_{c})-k_{\rm B}T\Big\{ 2 \sum_{d}n_{d}b_{cd}
\nonumber\\
+ 3 \sum_{de}n_{d}n_{e}b_{cde}+...\Big\}~,
\end{eqnarray}
where $p_{\rm id}(T,V,N_{c}) =k_{B}T\sum_{c}\frac{N_{c}}{V}$
and $\mu_{c,\rm id}(T,V,N_{c})=k_{B}T\ln\left(\frac{\Lambda_{c}^{3}N_{c}}{V g_c}\right)+E^{(0)}_c$
are the ideal parts of the pressure and the chemical potential, respectively (for the hydrogen atom the degeneracy factor is $g_a=4$). 
Note that in relativistic approaches the chemical potentials are gauged including
the rest mass of the constituents, as discussed for the Hamiltonian (\ref{eq:2}).

Working with the grand canonical ensemble, it is convenient to introduce the fugacities $z_c=e^{\beta (\mu_c-E_c^{(0)})}$ and to expand the
grand canonical potential with respect to the fugacities \cite{huang}, see also App.~\ref{app:fugacExp},
\begin{eqnarray}
J = - p V =-k_{\rm B}T V\Bigg[\sum_c \frac{g_c}{\Lambda_c^3}\Big(z_c 
\nonumber\\
+\sum_d\tilde{b}_{cd} z_c z_d + \dots \Big) \Bigg],
\end{eqnarray}
where $\tilde{b}_{cd}=b_{cd}g_d/\Lambda_d^3$ are the dimensionless second virial coefficients.
In the low density case $n_c\Lambda_c^{3}\ll1$, the relation $z_{c,\rm id}=n_c\Lambda_c^{3}/g_c$ results
for the ideal system. Since the different ensembles are equivalent in the thermodynamic limit, the fugacity expansion is shown
to be equivalent to the expansion with respect to the particle densities after substitution of the variables; for details see Ref.\cite{huang}.

Chemical equilibrium for a reaction $\nu_a A + \nu_b B \leftrightharpoons \nu_c C$ between the components $A,B,C$
gives the relation $\nu_a \mu_a + \nu_b \mu_b = \nu_c \mu_c$. This way, the thermodynamic potentials finally depend only on
the total densities of the constituents, or their chemical potentials, since the total number of the constituent particles is conserved.
The densities of the composites or their chemical potentials can be eliminated, using a mass action law or a Saha equation.

Note that it is possible to derive the cluster virial expansion in a systematic way, starting from a quantum statistical approach \cite{R}.
The spectral function of the elementary particle propagators is related to the self-energy. Within a cluster decomposition of the
self-energy, the contribution of the different clusters can be identified considering partial sums of ladder diagrams. The first-principle approach
gives a consistent introduction of the chemical picture avoiding double countings, and allows for systematic improvements.

In this paper we are concerned with the evaluation of the second virial coefficient that describes the non-ideality corrections in lowest order with respect to the densities.
According to Beth and Uhlenbeck \cite{BU,kremp,landau}, for a central symmetric interaction potential the following formula has been derived, which relates the second virial coeffient to the energy eigenvalues
$E^{cd}_{n\ell}$ of the two-particle bound state $\left| n\ell \right.\rangle$
and the scattering phase-shifts $\eta^{cd}_\ell(E)$ describing the two-particle
scattering states,
% \begin{widetext}
% \begin{equation}
%  \label{eq:7}
%  b_{cd}=-\frac{\Lambda_d^3}{2s_d+1}\left(\delta_{cd}\tilde{b}_{2}^{(0)}+2\sqrt{2}\,\sum_{\ell=0}^{\infty}(2\ell+1)
%  \left[ 1 \pm \delta_{cd} \frac{(-1)^\ell}{2s_c+1} \right]
%  \bigg[ \sum_{n\geq \ell+1} \mathrm{e}^{-\beta E^{cd}_{n\ell}} +\frac{\beta}{\pi}\int_{0}^{\infty}\mathrm{e}^{-\beta E}\,\eta_\ell^{cd}(E)\,dE
%  \bigg]\right)~,
% \end{equation}
% \end{widetext}
%  where $\ell$ denotes the angular momentum of the two-particle system,
%  \begin{equation}
%    \tilde{b}_{2}^{(0)} = \left\{ \begin{array}{ll}
%      2^{-5/2} & \textrm{(ideal Bose gas)}\\
%      -2^{-5/2} & \textrm{(ideal Fermi\ gas)}\\
%    \end{array} \right.
%  \end{equation}
% is the second virial coefficient for the ideal quantum gas.

\begin{equation}
 \label{eq:7}
 b_{cd}=\frac{\Lambda_d^3}{g_d}\left[\delta_{cd}\tilde{b}_{cd}^{(0)}+\tilde{b}_{cd}^{\rm bound}+\tilde{b}_{cd}^{\rm sc}\right]~,
\end{equation}
 where 
 \begin{equation}
   \tilde{b}_{cd}^{(0)} = \left\{ \begin{array}{ll}
     2^{-5/2} & \textrm{(ideal Bose gas)}\\
     -2^{-5/2} & \textrm{(ideal Fermi\ gas)}\\
   \end{array} \right.
 \end{equation}
is the second virial coefficient for the ideal quantum gas. 

%\textbf{ist es jetzt richtig? wenn ja, dann mussen wir noch text veraendern}
In this work we calculate the second virial coefficient for the $e-a$ contribution. 
The spin degrees of freedom of the bound electron and the scattering electron give rise to a singlet
(antiparallel electron spins) and  a triplet (parallel electron spins) scattering state;
the proton spin contributes a spin degeneracy factor $g_i=2$. The total second virial 
coefficient $b_{ae}=b_{ea}$ is defined with the total density of atoms and electrons.
It is decomposed into the singlet contribution and the triplet contribution, so that
$b_{ae}=b_{ae}^{\rm singlet}+b_{ae}^{\rm triplet}$. For convenience, we introduce the
dimensionless coefficient $\tilde{b}$ that appears in the fugacity expansion as
$\tilde{b}_{ea}=\frac{g_a}{\Lambda_a^3}b_{ea}$, $\tilde{b}_{ae}=
\frac{g_e}{\Lambda_e^3}b_{ae}$. Note that $\tilde{b}_{ea}$ is no longer symmetric with respect to a change of
indices, instead, we find  $\tilde{b}_{ea}=2\Big(\frac{m_a}{m_e}\Big)^{3/2}\tilde{b}_{ae}$.
Since $m_e\ll m_a$ we have $\Lambda_{ea}^3/\Lambda_{a}^3\approx1$, from which follows that the dimensionless
second virial coefficient is again the sum of the corresponding dimensionless singlet and triplet
coefficients,
$\tilde{b}_{ae}=\tilde{b}_{ae}^{\rm singlet}+\tilde{b}_{ae}^{\rm triplet}$.

The bound part of the second virial coefficient for the singlet state $\tilde{b}_{ae}^{\rm bound, singlet}$
has the following form:     
 \begin{eqnarray}
  \label{eq:b}
 \tilde{b}_{ae}^{\rm bound, singlet}=\frac{1}{4}\sum_{\ell=0}^{\infty}(2\ell+1)
%  \Big[ 1 \pm \delta_{cd} \frac{(-1)^\ell}{2s_c+1} \Big]
\sum_{n\geq \ell+1} \mathrm{e}^{-\beta E^{ae}_{n\ell}}~,
 \end{eqnarray}
 where $\ell$ denotes the angular momentum of the two-particle system.
The scattering part of the second virial coefficient $\tilde{b}_{ae}^{\rm sc}$ consist of the singlet and triplet parts: 
\begin{widetext} 
\begin{eqnarray}
\tilde{b}_{ae}^{\rm sc, singlet}=\frac{1}{4}\sum_{\ell=0}^{\infty}(2\ell+1)
  %\Big[ 1 \pm \delta_{cd} \frac{(-1)^\ell}{2s_c+1} \Big]
  \bigg[\frac{1}{\pi}\int_{0}^{\infty}\mathrm{e}^{-\beta E}\,\frac{d}{dE}\eta_\ell^{ae,\rm singlet}(E)\,dE
  \bigg]~
 \end{eqnarray}
\begin{eqnarray}
\tilde{b}_{ae}^{\rm sc, triplet}=\frac{3}{4}\sum_{\ell=0}^{\infty}(2\ell+1)
  %\Big[ 1 \pm \delta_{cd} \frac{(-1)^\ell}{2s_c+1} \Big]
  \bigg[\frac{1}{\pi}\int_{0}^{\infty}\mathrm{e}^{-\beta E}\,\frac{d}{dE}\eta_\ell^{ae, \rm triplet}(E)\,dE
  \bigg]~ 
\end{eqnarray}
\end{widetext}
At this point, we would like to make a short note regarding different forms of the Beth-Uhlenbeck formula Eq.(\ref{eq:7}) that can
be found in the literature. We use formula Eq.(\ref{eq:7}), which has
been derived originally by Beth and Uhlenbeck \cite{BU}, see also Refs~\cite{huang,landau}.
After partial integration of the Eq.(\ref{eq:7}) one
obtains another form for the Beth-Uhlenbeck formula, see e.g. \cite{roepke_bu,horowitz},
where the scattering phase shift arises instead of its derivative, and from the integration an additional
term $-\eta_\ell^{cd}(0)/\pi$ appears, which is sometimes condensed into the bound part \cite{horowitz}.

Coming back to the partially ionized plasma, the virial expansion is diverging for the long-range Coulomb interaction. This refers to the $e-e,\, e-i,\,i-i$ contributions. Partial summations lead to convergent results, and the expansion of the thermodynamic potentials contains also terms with $n_c^{1/2}$ and $\log n_c$, see \cite{kraeft}.
The contribution of the scattering and bound parts of the second virial coeffient for atom-atom interaction
was calculated in Ref. \cite{schlanges2}. In this work, we evaluate the second virial coefficient Eq.~(\ref{eq:7}) for the electron-atom interaction.

%%%%%%%%%%%%%%%%%%%%%%%%%%%%%%%%%%%%%%%%%%%%%%%%%%%%%%%%%%%%%%%%%%%%%%%%%%%%%%%%%%%%%%%%%%%%%%%%%%
\section{Elastic electron-atom scattering and phase-shifts\label{sec:phaseshifts}}

\subsection{Experimental data and first-principle calculations}
 
\begin{figure}[htp]
\includegraphics[width=0.37\textwidth,angle=-90]{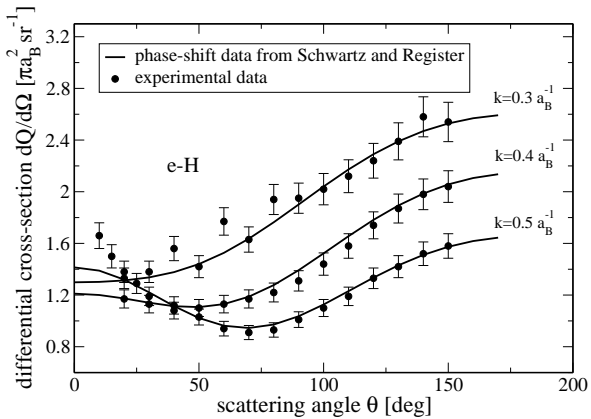}
\caption{Differential electron-atom ($e-a$) scattering cross-section as a function of the scattering angle for electron wavenumber
$k=0.3~a_{B}^{-1}$, $k=0.4~a_{B}^{-1}$, $k=0.5~a_{B}^{-1}$;
solid lines - phase-shift data from the variational method \cite{schwartz,register}, circles - experimental data \cite{willams1975}.
}
\label{fig:p1}
\end{figure}

\begin{figure}[tp]
\includegraphics[width=0.37\textwidth,angle=-90]{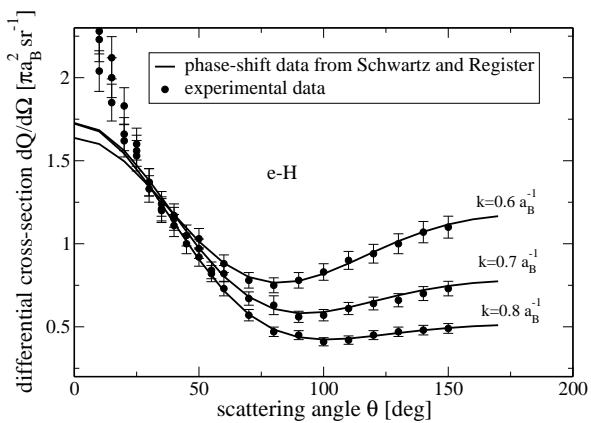}
\caption{Differential electron-atom ($e-a$) scattering cross-section as a function of the scattering angle for electron wavenumber $k=0.6~a_{B}^{-1}$, $k=0.7~a_{B}^{-1}$, $k=0.8~a_{B}^{-1}$.
\label{fig:p2}}
\end{figure}

The Beth-Uhlenbeck formula relates the second virial coefficient to few-body properties.
For the electron-atom contribution, the relevant quantities are
the phase-shifts for the elastic electron-atom scattering
as well as the possible bound state energies.
No direct measurements of the electron-atom scattering phase-shift are available in the
literature, only scattering cross-sections (i.e. the modulus of the phase-shift)
have been measured. Accurate data for the angular resolved scattering cross-section
were obtained by Williams {\it et al}. \cite{willams1975} for electron energies between $0.58\,\mathrm{eV}$ and
$8.7\,\mathrm{eV}$ and by Gilbody {\it et al}. \cite{gilbody} for $3.4\,\mathrm{eV}$,
see the review in Ref.~\cite{willams1998}. The data are shown in Figs. \ref{fig:p1} and \ref{fig:p2}.
A bound state H$^-$ is measured at energy $(0.754\pm 0.002)\,\textrm{eV}$ \cite{culloh}.
In this measurement,
the electron affinity was obtained via the threshold energy for photodissociative formation
of ion pairs from H$_2$ \cite{culloh}. The ion pair threshold was combined with the ionization potential of the
hydrogen atom and the bound dissociation energy of H$_2$ to obtain a lower bound to
the electron affinity.
This results is in agreement with the theoretical value
of $0.75421\,\text{eV}$ reported by Pekeris \cite{pekeris2} for the singlet bound state
of H$^-$.

Theoretical calculations for the $e-a$ scattering phase-shifts are abundant. The spins of the two electrons are combined to a singlet or triplet state, whereas the orbitals are determined by a three-body Schr\"odinger equation and the symmetry condition for fermions. Frequently used methods are the
close-coupling approximation \cite{burke}, the R-matrix method \cite{scholz},
direct numerical solution of the Schr\"odinger equation \cite{wang},
the wave expansion method \cite{calogero,babikov}, and variational calculations. Using the variational method, Schwartz {\it et al}. \cite{schwartz} obtained the phase-shifts for the $s$-waves (orbital momentum $\ell = 0$ of the $e - a$ system)
in the singlet and triplet channels.
For higher orbital moments $\ell=1,2$ calculations have been performed by Armstaed \cite{armstead} and Register \cite{register}.
 Comparing the experimental data with results of numerous theoretical approaches, see Refs~\cite{willams1975,register,faasen}, it was concluded that the variational approach is the most reliable and most accurate
method in reproducing the experimental scattering cross-sections.

In Figs.~\ref{fig:p1} and \ref{fig:p2} we show the differential cross-section as calculated from the phase-shifts given in Refs~\cite{schwartz,register} compared to the experimental data from Ref.~\cite{willams1975}.
Good agreement between theory and experiment is observed for scattering angles larger than $\theta\simeq 25^\circ$.
Deviations below $25^\circ$ are due to the neglect of higher orbital momenta $\ell>2$ in the variational calculations \cite{willams1975}.

\subsection{Polarization potential models}

Before presenting results for the second $e-a$ virial coefficient using the data given above,
we discuss the introduction of pseudopotentials.
One of the advantages of introducing pseudopotentials is the possibility to describe medium
effects such as screening and quantum degeneracy.
Polarization effects at long distances and exchange effects at short distances play a key role in the
$e-a$ interaction in a dense plasma. Below we use polarization potential models, which include
these many-body
features to calculate the scattering phase-shifts. Of course, the introduction of
a local and instant pseudopotential for the $e-a$ interaction is only an approximation. A
rigorous treatment involves the
three-particle T matrix, which is non-local due to exchange effects and depends on energy. We
do not describe this problem in detail here.

We calculate the phase-shifts on the basis of the wave expansion method \cite{babikov}.
This approach yields the phase-shift as the solution to the so-called Calogero \cite{calogero} equation,
\begin{eqnarray}
\label{eq:10}
\frac{d\eta_{\ell}^{cd}(k,r)} {dr}=-\frac{1}{k}U(r)\Big[\cos\eta_{\ell}^{cd}(k,r)J_{\ell}(k,r)
\nonumber\\
-\sin\eta_{\ell}^{cd}(k,r)n_{\ell}(k,r)\Big]^2
\end{eqnarray}
 which is a first order non-linear differential equation.
 Here, $J_{\ell}(k,r)$ and $n_{\ell}(k,r)$ are the Rikkati-Bessel functions \cite{AbramowitzStegun}, $U(r)=(2m_{r}/\hbar^2)V_{cd}(r)$,
 $V_{cd}(r)$ is the interaction potential and $m_{r}$ is the reduced mass of the two-particle system $c-d$.
As initial condition we apply $\eta_{\ell}^{cd}(k,0)=0$, thereby fixing the arbitrary phase-offset.
The phase-shifts are defined by
$\eta_{\ell}^{cd}(k)=\lim_{r\to\infty}\eta_{\ell}^{cd}(k,r)$ and depend only on the wavenumber $k$.
As its main advantage, the wave expansion method allows to study directly the influence of the interaction potential on the phase-shifts.
Secondly, equation (\ref{eq:10}) is easier to solve than the full Schr\"odinger equation applying, e.g., variational methods.

As the electron-atom interaction potential we consider the Buckingham model \cite{rrz}
\begin{equation}
\label{eq:11}
V_{ea}(r)=-\frac{\alpha e^2}{2(r^2+r_{0}^2)^2}~,
\end{equation}
$\alpha=4.5\, a_{\rm B}^3$ is the hydrogen atom polarizability (here and henceforth we measure distances
in units of the Bohr radius, $a_\mathrm{B}=0.529\,\text{\AA}$). The cutoff radius $r_0$ is used to regularize the behaviour at small distances. Its value
given in the literature \cite{redmer1997} is $r_0=1.4565\,a_{\rm B}$. However, we suggest here the use of a different value $r_0=1.033\,a_\mathrm{B}$, which yields the correct H$^{-}$ ion ground state energy $E_{0} = -0.7542\,\mathrm{eV}$ as the eigenvalue of the effective radial Schr\"odinger equation \cite{kraeft}.
At large distances $r\gg r_0$ the Buckingham potential describes the typical $1/r^4$ behaviour of the electron potential energy in the field of the
polarizable H-atom.
 
Secondly, we employ the
polarization potential model that was suggested for semiclassical plasmas in Ref. \cite{RDO}:
\begin{eqnarray}
 \label{eq:12b}
V_{ea}(r)=-\frac{e^{2}\alpha}{2r^{4}(1 -4\lambda_e^2 / r_D ^2)}\times
\Big(e^{ -Br}(1 + Br) 
\nonumber\\
- e^{ - Ar}(1 + Ar)\Big)^{2},
\end{eqnarray}
 where $A^2=(1+\sqrt{1-4\lambda_e^2/r_D^2})/(2\lambda_e^2)$,
$B^2=(1-\sqrt{1-4\lambda_e^2/r_{D}^2})/(2\lambda_e^2)$,
and $\lambda _{e} =\Lambda_e/2\pi$ is the electron thermal de Broglie wave-length, $r_D=[\epsilon_0\,k_\mathrm{B}T/n_\mathrm{e}^\mathrm{free}e^2]^{1/2}$
is the Debye radius due to
electrons. This model depends on two parameters $\lambda_e$ and $r_D$.
At large distances the RDO potential is weaker than the Buckingham model (\ref{eq:11}) due to
screening. The strength of the model at short distances is given by $-\alpha e^2/8\lambda_e^4$.
Fixing $r_{D} = 4.84\,a_{\rm B}$ and $\lambda_e = 0.62\,a_{\rm B}$,
the H$^{-}$ ground state energy is found at the correct energy. 

At this point we want to remark that the bound state occurs only in the singlet scattering channel,
due to the strongly repulsive exchange interaction in the triplet channel. However, neither the Buckingham model, Eq.~(\ref{eq:11}),
nor the RDO model, Eq.~(\ref{eq:12b}), take into account this exchange effect. A convenient method to overcome this problem consists in using a separable potential \cite{Mongan,yamaguchi}. This will be discussed in the following subsection \ref{subsec:sep}.

\subsection{Separable potential method \label{subsec:sep}}

Separable potentials have been used extensively in nuclear physics to parametrize the
nucleon-nucleon interaction \cite{Mongan}.
It can be shown that any interaction potential can be approximated by a sum of separable potentials \cite{EST}.

 We characterize different channels by spin and angular momentum and assume that there is no coupling between different channels. We consider a rank-two separable
potential in momentum representation to describe attraction at long distances and repulsion at short distances,
\begin{equation}
V(p,p')=\lambda_1w_1(p)w_1(p')+\lambda_2w_2(p)w_2(p')
\label{eq:s1}
\end{equation}
where $w_1(p)$, $w_2(p)$ are Gaussian form factors $w_i(p)=\exp{(-p^2/b_i^2)}$, and $\lambda_i$, $b_i$ are the strength and interaction range, respectively. We determine these parameters by fitting the
phase-shifts obtained from the separable potential to the experimental data. Thereby we exploit the definition of the phase-shifts
as the argument of the T-matrix,
\begin{equation}
\tan\eta(p)=\frac{\Im{T(p,p',\frac{p^2}{2m_{r}})}}{\Re{T(p,p',\frac{p^2}{2m_{r}})}},
\label{eq:17b}
\end{equation}
The calculation of the T-matrix for the separable potential, which amounts to the solution of the effective Schr\"odinger equation
\begin{equation}
 \frac{p^2}{2m_{r}}\psi(p)+\sum_{p'}V(p,p')\psi(p')=E_0\psi(p)~,
\label{eq:s3}
\end{equation}
is detailed in the App. \ref{app:Tmatrix}.

Fig. \ref{fig:1}
shows the phase-shifts for the singlet and triplet
scattering channels obtained by the separable potential in comparison with data of Ref. \cite{schwartz}. The best fit parameters are summarized in Tab. \ref{tabs2}.
Column 4, 5 and 6 give the scattering length $a$ and the effective range $R$ from the effective radius theory, and the binding energy $E_0$ used to fix two of the parameters $\lambda_i$ and $b_i$, the remaining two being fixed directly by comparison to the experimental phase-shifts.
The effective radius theory was applied e.g. in
Ref. \cite{starostin} to calculate the influence of atomic and molecular contributions to the EOS of hydrogen plasma. %using the simple effective radius
However, the effective radius theory is limited to $s$-wave scattering, and therefore to low energies. Furthermore, in Ref. \cite{starostin}, the spin dependence of the $e-a$ interaction is neglected.
\begin{table}[htp]
\center
\caption{\label{tabs2}Parameters of the rank-two and rank-one separable potentials ($\lambda_1,\lambda_2$, in Hartree, and $b_1,b_2$ in $a_B$), the scattering length ($a$), the effective range ($R$), the binding energy ($E_0$) for the singlet ($S=0$) and the triplet ($S=1$) channels of $s$-wave.}
\begin{ruledtabular}
\item[]
\begin{tabular}{c|ccccccc}
& $\lambda_1$ & $\lambda_2$&$b_1$&$b_2$&$a/a_{\rm B}$&$R/a_{\rm B}$&$E_0$, Hartree\\
\hline
\multicolumn{8}{c} {rank-two separable potential}\\
\hline
 $S=0$ &$-25.4$ & $10$ &$0.8$&$0.787$& $5.965$ &$-$& $-0.06899$\\
 $S=1$ &$37$ &$40$ &$0.8$&$0.787$& $1.97$&$-$ &$-$\\
\hline
\multicolumn{8}{c} {rank-one separable potential}\\
\hline
 $S=0$ &$-45.4$&0 & $0.4705$ &0 & $5.965$ & $3.32$&$-0.0474$\\
 $S=1$ &$77.67$ &0 & $0.9$ &0 & $1.77$ & $1.1$&$-$\\
\end{tabular}
\end{ruledtabular}
\end{table}

Choosing $\lambda_2=0$ and $b_2=0$ in Eq.~(\ref{eq:s1}), a rank-one separable potential is obtained.
Parameters of a rank-one separable potential are given in the Tab. \ref{tabs2} (bottom part).
Using the properties of the T-matrix (see Refs. \cite{Mongan,kraeft}),
we found the binding energy $E_0$ of the $H^{-}$ ion for both versions of the
separable potentials, given in the last column of Tab. \ref{tabs2}.
The experimental value of the binding energy
is $E_0=-0.0277$ Hartree ($-0.7542\,\text{eV}$) \cite{culloh}. 
The two-particle properties, i.e. scattering phase-shifts and the bound state properties, can be reproduced in certain approximation by separable 
potentials. We expect that increasing the rank of the potential, the experimental values for the two-particle properties are better realized.

\subsection{Phase shifts data}
\begin{figure}[htp]
\includegraphics[width=0.37\textwidth,angle=-90]{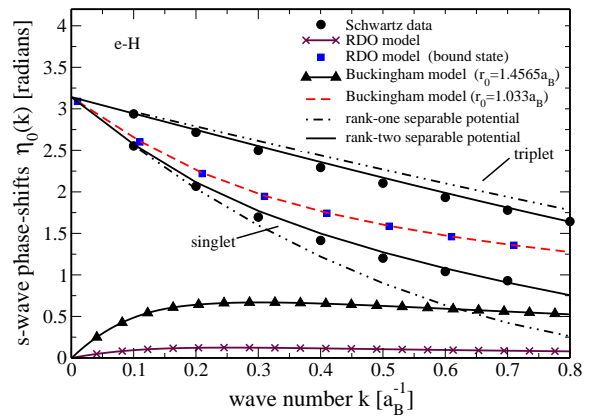}
\caption{Electron-atom (e-H) scattering phase-shifts
 as a function of wave number $k$ for $\ell=0$. Shown are experimental data of Schwartz \cite{schwartz} for the singlet and triplet channels, model calculations using the Buckingham [Eq.(\ref{eq:11})] and the RDO pseudopotential [Eq.(\ref{eq:12b})] at different parameters, as well as different separable potentials.}
\label{fig:1}
\end{figure}
\begin{figure}[htp]
\includegraphics[width=0.37\textwidth,angle=-90]{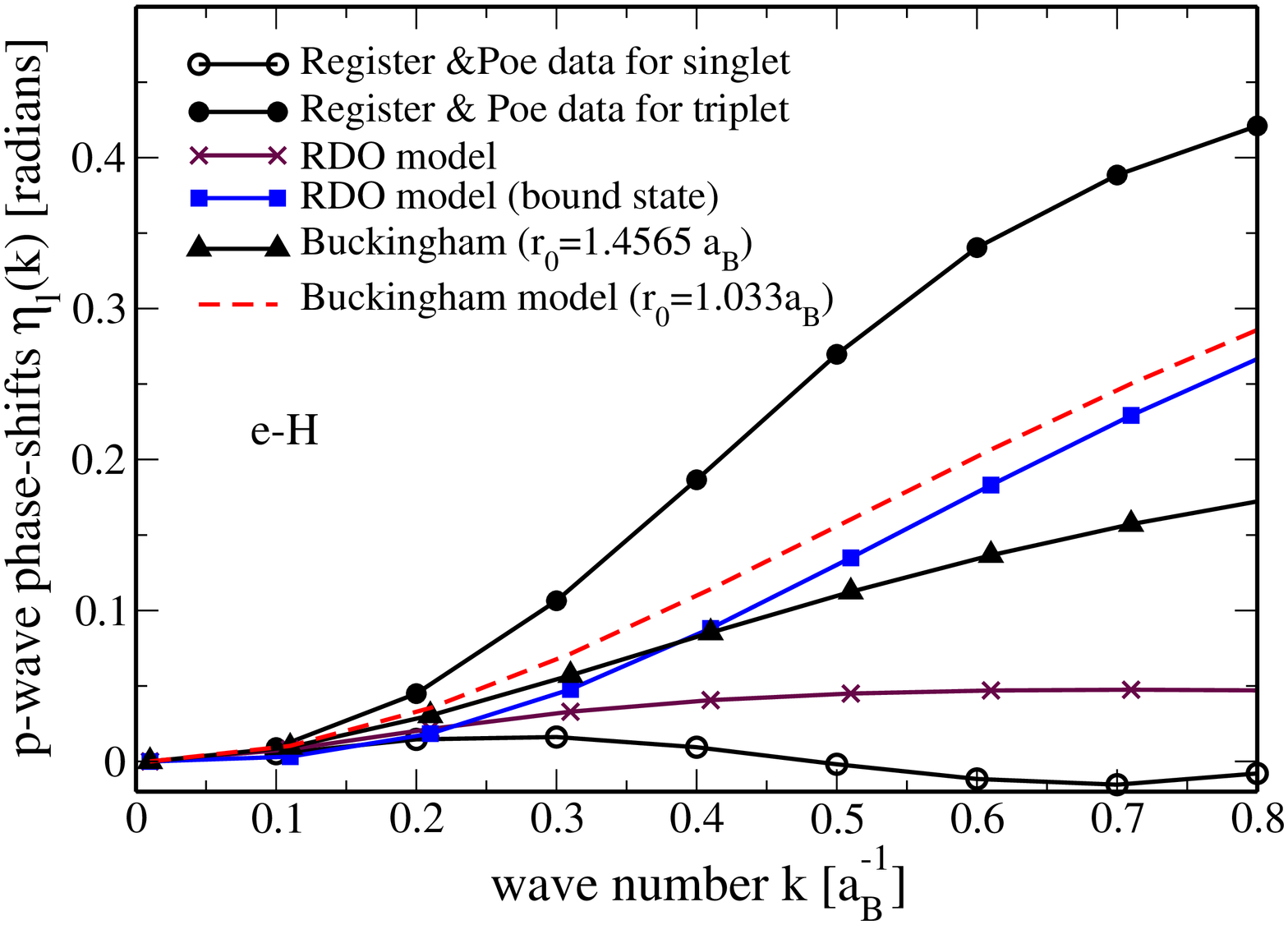}
\caption{Electron-atom (e-H) scattering phase-shifts as a function of wave number $k$ for $\ell=1$.}
\label{fig:2}
\end{figure}
\begin{figure}[htp]
\includegraphics[width=0.37\textwidth,angle=-90]{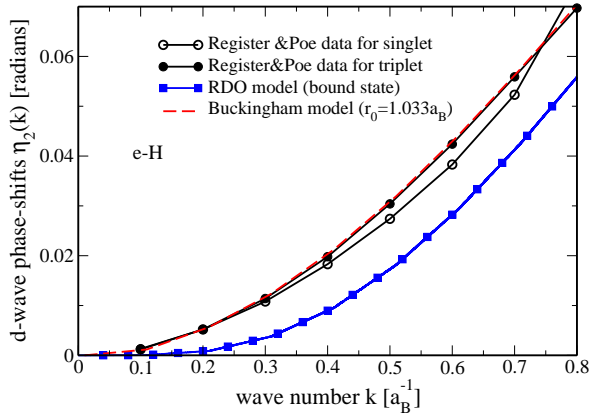}
\caption{Electron-atom (e-H) scattering phase-shifts as a function of wave number $k$ for $\ell=2$.}
\label{fig:3}
\end{figure}
Using the potentials (\ref{eq:11}) and (\ref{eq:12b}), 
the Calogero equation (\ref{eq:10}) is solved.
The $s$-wave scattering phase-shifts obtained in this way
are plotted as a function of the wavenumber $k$ in Fig. \ref{fig:1}.
We compare our calculations
to the experimentally
validated data by Schwartz {\it et al}., employing different choices for
the cutoff parameter $r_0$ in the case of the Buckingham potential, and $r_D$ and $\lambda_e$
for the RDO potential.

At $k=0$, the singlet phase-shifts from \cite{schwartz} tend to $\eta_0=\pi$,
corresponding to one bound state as follows from
the classical Levinson theorem \cite{levinson}, $\eta(0)=n\pi$ ($n$ is number of bound states).
The polarization potential method gives a bound state if the screening parameters
that reproduce the
correct H$^{-}$ binding energy are applied
(red dashed curve for Buckingham potential, $r_0=1.033\,a_\mathrm{B}$ and blue square curve for RDO
potential, $r_{D} = 4.84\,a_{\rm B}$ and $\lambda_e = 0.62\,a_{\rm B}$). Taking the
original screening parameters in both models, we find vanishing phase-shifts at
$k=0$, i.e. no bound state.
The solid line and dash-dotted line correspond to phase-shifts for rank-two and rank-one
separable potential, respectively. The phase-shifts from the rank-two separable potential
fully coincide with the experiments, whereas the results for the rank-one separable potential
deviate
at high values of $k$ (respectively $E$).

We consider the $s$-wave scattering phase-shifts in the triplet
channel in
Fig. \ref{fig:1}. At zero-energy the phase-shift starts off at
$\pi$. Having in mind that the effective interaction between the
electron and atom in the triplet channel is repulsive, this behavior
obviously contradicts Levinson's theorem that predicts the scattering
phase-shift to increase by $\pi$ with every occurring bound state. To
resolve this inconsistency,
the classical Levinson theorem for one-body problems has
been generalized for the scattering on a compound target by Rosenberg and Spruch
\cite{rosenberg_spruch1}. The generalized Levinson
theorem states that the phase-shift at vanishing energy is
$\eta(0)=(n_{\rm Pauli}+n)\pi$, where $n_{\rm Pauli}$ is the
number of states from which the particle is excluded by
the Pauli principle. $n_{\rm Pauli}$ is defined by the number
of nodes in the one-particle wave function. Application of the
generalized Levinson theorem to the electron-hydrogen
triplet scattering shows the one-particle wave function
has one node, that means the zero-energy triplet phase
shift is a nonzero multiple of $\pi$
\cite{rosenberg_spruch2}. Since the triplet
electron-hydrogen wave function is spatially
antisymmetric, the equivalent one-body wave function is
orthogonal to the hydrogenic ground-state function and
must have at least one node. Thus, our result for the behavior at
zero-energy
of the triplet phase-shift does not predict a
triplet bound state $H^{-}$. This
agrees with previous investigations of scattering of electrons on hydrogen atoms at low energies, where the scattering length
was defined under the assumption that the negative hydrogen ion
$H^{-}$ can be formed only in the singlet channel \cite{ohmura,geltman,omalley1,omalley2,omalley3}.
In our calculations of the second virial coefficient we
consider only the singlet bound state of $H^{-}$.

Figs. \ref{fig:2}
and \ref{fig:3} show the calculated phase-shifts for $p$ and $d$ waves, respectively. For comparison, the data from variational
calculations \cite{register} are also presented. The phase-shift is zero (no bound states)
at the origin and increases monotonically. The phase-shifts for $\ell=1$ and $\ell=2$ are very small in comparison to the
$s$-wave data for the low-energy range. In general, terms from higher orbital moments are negligible for low-energies. 
Hence, we only apply phase-shift data for $\ell=0,1,2$ in further calculations.
\subsection{Results for the second virial coefficient\label{sec:2ndvc}}

The good agreement between experimental cross-sections and the variational scattering phase-shifts from Refs.~\cite{schwartz,register} allows us to use the latter as ``experimentally confirmed'' data
for calculations of the second virial coefficient using the Beth-Uhlenbeck formula (\ref{eq:7}). The phase-shifts obtained from pseudopotential models (\ref{eq:11}) and (\ref{eq:12b}) are not applied here to calculate $\tilde{b}_{ae}$.

Tabs.~\ref{tab2} and \ref{tab3} show results for the normalized second virial coefficients
$\tilde{b}_{ae}^{\rm sc,singlet}(T)$, $\tilde{b}_{ae}^{\rm bound,singlet}(T)$ for the singlet channel and
$\tilde{b}_{ae}^{\rm sc,triplet}(T)$ for the triplet channel, respectively.
The second, third and fourth columns of both tables present data for the contribution of $s$, $p$ and $d$ waves to the scattering part of the second virial coefficient for singlet and triplet channels, respectively. Higher order contributions are small and negligible for this temperature range. The singlet bound part of the second virial coefficient for the singlet channel is shown as the fifth column of Tab. \ref{tab2}. The binding energy is taken as $E^{eH}_{B}=-0.7542$ eV \cite{culloh}. The full scattering part of the second virial coefficient is shown in the sixth column of both tables. The last column of the Tab. \ref{tab3} presents the results for the full second virial coefficient: $\tilde{b}_{ae}(T)=\tilde{b}_{ae}^{\rm sc,singlet}(T)+\tilde{b}_{ae}^{\rm sc,triplet}(T)+\tilde{b}_{ae}^{\rm bound, singlet}(T)$. 

We find that the scattering contributions to the second virial coefficient increases with temperature in contrast to the
bound state contribution. In Tabs. \ref{tab2} and \ref{tab3} the $s$-wave contribution to the second virial coefficient is the dominant term, $p$-wave and
$d$-wave contributions are of the order of few percent.

\begin{table*}[htp]
\center
\caption{\label{tab2}The singlet scattering and bound parts of the second virial coefficient for $e-H$ interaction. Contribution of 
different partial waves and bound state contribution are given.}
\begin{ruledtabular}
%\item[]
\begin{tabular}{c|c|c|c|c|c}
$T$, K & $\tilde{b}^{\rm sc,singlet}_{ae}$, $s$-wave&$\tilde{b}^{\rm sc,singlet}_{ae}$, $p$-wave&$\tilde{b}^{\rm sc,singlet}_{ae}$,
$d$-wave&$\tilde{b}^{\rm bound, singlet}_{ae}$&$\tilde{b}^{\rm singlet}_{ae}$ full\\
\hline
5000& -0.0499& 0.0012&   0.0007&   1.4401&   1.3922\\

6000& -0.0541& 0.0015&   0.0008&   1.0756&   1.0239\\

7000& -0.0578& 0.0017&   0.0010&   0.8732&   0.8181\\

8000& -0.0611& 0.0018&   0.0012&   0.7468&   0.6887\\

9000& -0.0642& 0.0020&   0.0013&   0.6613&   0.6005\\

10000&-0.0670& 0.0021&   0.0015&   0.6000&   0.5366\\

11000&-0.0696& 0.0022&   0.0016&   0.5541&   0.4883\\

12000&-0.0720& 0.0022&   0.0018&   0.5185&   0.4505\\

13000&-0.0743& 0.0023&   0.0019&   0.4902&   0.4202\\

14000&-0.0765& 0.0023&   0.0021&   0.4672&   0.3952\\

15000&-0.0785& 0.0024&   0.0022&   0.4481&   0.3742\\

20000&-0.0873& 0.0024&   0.0029&   0.3873&   0.3054\\

30000&-0.1004& 0.0022&   0.0043&   0.3347&   0.2409\\

40000&-0.1099& 0.0019&   0.0057&   0.3111&   0.2089\\

50000&-0.1172& 0.0016&   0.0072&   0.2978&   0.1894\\

60000&-0.1230& 0.0015&   0.0088&   0.2892&   0.1765\\

70000&-0.1276& 0.0015&   0.0105&   0.2833&   0.1678\\

80000&-0.1312& 0.0017&   0.0125&   0.2789&   0.1620\\

90000&-0.1340& 0.0022&   0.0148&   0.2755&   0.1584\\

100000&-0.1361& 0.0028&   0.0172&   0.2728&   0.1568\\
\end{tabular}
\end{ruledtabular}
\end{table*}
\begin{table*}[htp]
\center
\caption{\label{tab3}The triplet scattering part of the second virial coefficient for $e-H$ interaction. Contribution of different 
partial waves, last coloumn: full second virial coefficient $\tilde{b}_{ae}$. }
\begin{ruledtabular}
\item[]
\begin{tabular}{c|c|c|c|c|c}
$T$, K & $\tilde{b}^{\rm sc,triplet}_{ae}$, $s$-wave&$\tilde{b}^{\rm sc,triplet}_{ae}$, $p$-wave&$\tilde{b}^{\rm sc,triplet}_{ae}$,
$d$-wave&$\tilde{b}^{\rm triplet}_{ae}$ full&$\tilde{b}_{ae}$ full\\
\hline

5000&   -0.0548&   0.0121&   0.0024&   -0.0402&   1.3519\\

6000&   -0.0604&   0.0147&   0.0029&   -0.0426&   0.9812\\

7000&   -0.0654&   0.0174&   0.0033&   -0.0446&   0.7735\\
8000&   -0.0702&   0.0200&   0.0038&   -0.0462&   0.6425\\
9000&   -0.0746&   0.0227&   0.0043&   -0.0475&   0.5529\\

10000&   -0.0788&   0.0254&   0.0048&   -0.0485&   0.4880\\

11000&   -0.0828&   0.0281&   0.0053&   -0.0494&   0.4389\\

12000&   -0.0866&   0.0307&   0.0057&   -0.0501&   0.4004\\

13000&   -0.0902&   0.0333&   0.0062&   -0.0506&   0.3695\\

14000&   -0.0937&   0.0360&   0.0067&   -0.0510&   0.3442\\

15000&   -0.0970&   0.0386&   0.0071&   -0.0512&   0.3230\\

20000&   -0.1121&   0.0512&   0.0094&   -0.0514&   0.2539\\

30000&   -0.1367&   0.0742&   0.0139&   -0.0485&   0.1924\\

40000&   -0.1567&   0.0943&   0.0183&   -0.0440&   0.1648\\

50000&   -0.1735&   0.1117&   0.0224&   -0.0393&   0.1501\\

60000&   -0.1882&   0.1270&   0.0264&   -0.0347&   0.1418\\

70000&   -0.2012&   0.1405&   0.0302&   -0.0304&   0.1373\\

80000&   -0.2128&   0.1525&   0.0338&   -0.0264&   0.1355\\

90000&   -0.2233&   0.1635&   0.0372&   -0.0226&   0.1358\\

100000&   -0.2328&   0.1735&   0.0403&   -0.0189&   0.1379\\

\end{tabular}
\end{ruledtabular}
\end{table*}

\section{Equation of State and Thermodynamics\label{sec:thermodynamics}}

\subsection{Composition}

The virial expansion allows to determine a thermodynamic potential that gives all thermodynamic variables. We discussed the free energy  $F(T, V, N_c)$ or the grand potential $-pV = J(T,V, \mu_c)$. 
Because of reactions in the system, the particle numbers of the different components are related by the chemical equilibrium conditions so that the number of independent variables is reduced. 
In the case of a hydrogen plasma considered here, we start from the particle numbers of free electrons, free ions, and atoms, disregarding heavier clusters. The atomic density is related to the
free electron and ion density due to the Saha equation  that follows from the equilibrium condition $ \mu_e+\mu_i=\mu_a$. The remaining two particle numbers, $N_c^{\rm tot}=N_c+N_a$ with $c=e,i$,
will coincide if a charge-neutral plasma is considered so that we end up with only one particle number $N=N_e^{\rm tot}=N_i^{\rm tot}$ for a charge-neutral hydrogen plasma in chemical equilibrium.

To derive the composition from the chemical equilibrium condition we express the chemical potentials in terms of the densities, see Eq. (\ref{eq:5}). In lowest order of the cluster virial expansion, the ideal Saha equation
\begin{equation}
\label{eq:saha}
 \frac{1-\alpha}{\alpha^2}=n_{e}^{\textrm{tot}}\Lambda_e^3\,\exp\left[\beta I^{\rm eff}(n_{e},T)\right]~,
\end{equation}
is obtained for the  degree of ionization $\alpha=n_e/n_e^\textrm{tot}$, where $I_{\rm id}^{\rm eff}(n_{e},T)=|E_{a}^{0}|$.

We will not discuss here the more general expressions where the excited states and higher clusters are included \cite{kraeft}. The thermal wavelength for the atom was approximated by the thermal wavelength for the ion.

Taking the non-ideal terms into account, e.g. according to a virial expansion, the composition follows from a Saha equation with shifted energies  \cite{kremp}
\begin{equation}
I^{\rm eff}(n^{\rm tot}_{e},n^{\rm tot}_{i},T)=|E_{a}^{0}|-\Delta_{a}+\Delta_{e}+\Delta_{i}~.
\end{equation}
The energy shifts $\Delta_c$ of the different components can be obtained from density expansions.
 As an approximation we take the Debye shift
$\Delta_{e}=\Delta_{i}=-\kappa e^2/2$ due to the Coulomb interaction between the charged particles,
$\kappa= [n_\mathrm{e}^\mathrm{free}e^2/\epsilon_0\,k_\mathrm{B}T]^{1/2}$ is the inverse Debye screening length. These terms are of the order $n_e^{1/2}$. The bound
energy shift  $\Delta_{a}$  is not taken into account here because it is of higher order in density. 

\begin{figure}[htp]
 \includegraphics[width=0.37\textwidth,angle=-90]{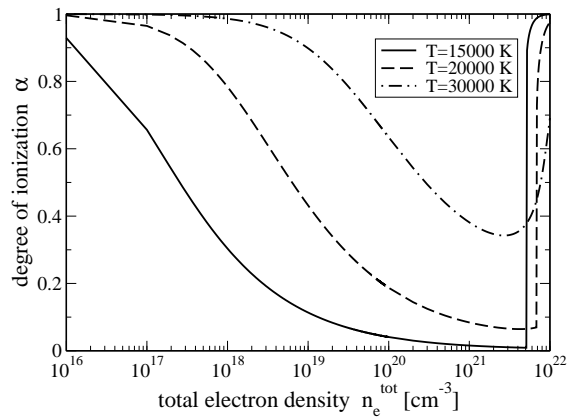}
\caption{Degree of ionization as a function of the total electron density for three different temperatures,
$T=1.5, 2, 3\cdot 10^{4}\,\text{K}$.
\label{figeq}}
\end{figure}

In Fig. \ref{figeq} we plot the solution of the Saha equation (\ref{eq:saha}) in dependence of the 
total electron density for temperatures $T=15000\,\text{K}, 20000\,\text{K}$, and $30000\,\text{K}$. The degree of ionization is decreasing with increasing 
density due to formation of bound states. 
The effective bound state ionization energy $I^{\rm eff}$ 
is lowering due to plasma screening. Ultimately, this leads to the Mott effect, i.e. the non-thermal
ionization at high densities, due to the lowering of the ionization threshold, 
leads to the abrupt increase of the ionization degree, see also
Refs.~\cite{kraeft,kremp}. We refrain from giving an exhaustive 
description of the Mott effect including more
sophisticated analysis of the shifts and restrict ourselves only to the general behavior of the
ionization degree. Note, that the virial expansion can only be applied to the low density range where the corrections are small.

\subsection{Chemical potential, free energy, pressure}

To discuss the contribution of electron-atom scattering to the
chemical potential (\ref{eq:5}), free energy (\ref{eq:3}), and pressure (\ref{eq:4})
we rewrite the definitions as
\begin{eqnarray}
 \Delta\mu_{ea}=\mu^e(T,n_c)-\mu_{\rm id}^e=-2k_B Tn_a b_{ea}(T)~,\\
\Delta F_{ea}=F(T,n_c)-F_{\rm id}
=p(T,n_c)V-p_{\rm id}V&&
\nonumber\\
=-2k_BTVn_en_ab_{ea}(T)~.&&
\end{eqnarray}
The remaining contributions to the second virial coefficient due to the other combinations
of components will not be discussed here, see \cite{kraeft,kremp}.

\begin{figure}[htp]
 \includegraphics[width=0.37\textwidth,angle=-90]{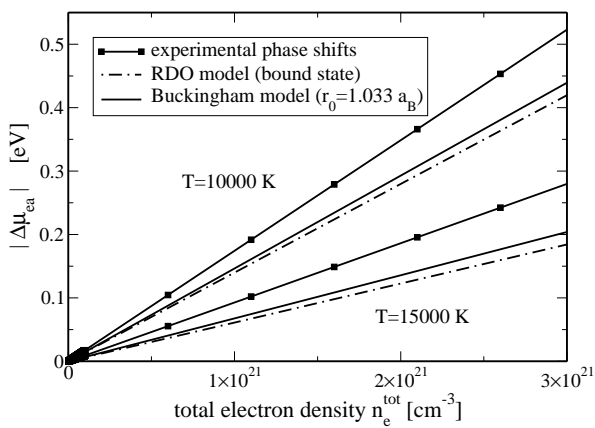}
\caption{Contribution of  electron--hydrogen interaction to the chemical potential as a
function of the total electron density; squares represent calculations based on
experimental phase-shift data, the solid line uses the Buckingham pseudopotential, the
dot-dashed line uses the RDO model.
\label{fig:deltach}}
\end{figure}

As mentioned before, the second virial coefficient can be decomposed into the singlet and triplet channel
and it is given as the sum of
scattering and bound state contributions,
\begin{eqnarray}
b_{ea}(T)=b_{ae}(T)=\frac{\Lambda_e^3}{2}\Big[\tilde{b}_{ae}^{\rm sc,singlet}(T)+\tilde{b}_{ae}^{\rm sc,triplet}(T)
\nonumber\\
+\tilde{b}_{ae}^{\rm bound, singlet}(T)\Big]~.
\end{eqnarray}
The various terms are given in Tabs~\ref{tab2} and \ref{tab3},

In Fig.~\ref{fig:deltach}, we plot the $e-$H scattering contribution to the chemical potential
$\Delta\mu_{ea}=-2k_B Tn_a b_{ea}(T)$ as a function of the total electron number density
for $T=10000\,\text{K}$ and $T=15000\,\text{K}$. In a similar way, we treat the $e-$H contribution to the free energy $\Delta F_{ea}=-2k_BTVn_en_ab_{ea}(T)~$ and to the pressure $\Delta p_{ea}=-2k_BT n_en_ab_{ea}(T)$. 

\subsection{Comparison of the results with the excluded volume concept} 

An alternative approach to evaluate the non-ideality term due to the neutral atoms is the excluded volume concept \cite{EbFortov}.
The excluded volume is defined by the filling parameter $\eta=\frac{4}{3}\pi r_a^3n_a$ as the volume that is occupied
by atoms such that the effective volume available for the moving particles is $V^{*}=V(1-\eta)$. 
The atom radius  $r_a$ is an empirical parameter of the order of $a_{\rm B}$ 
that has been fixed in different ways (see Ref. \cite{EbFortov}). Within the excluded volume concept, 
the non-ideality part of the free energy reads
\begin{eqnarray}
\label{eq:free1}
 \Delta F_{ea}^{\rm ex}&=&F_{\rm id}(T,V^{*},N_e,N_i,N_a)-F_{\rm id}(T,V,N_e,N_i,N_a)
\nonumber\\
&=&\frac{4}{3}\pi
 r_a^3k_{B}TN_a[n_e+n_i+n_a].
\end{eqnarray}
The corresponding second virial coefficient for the electron-atom pair results as
\begin{equation}
\label{eq:bex}
 b_{ea}^{\rm ex}=-\frac{2}{3}\pi r_a^3~. 
\end{equation}

It is instructive to note that this expression is equal to
the classical second virial coefficient within the Beth-Uhlenbeck approach using a hard-sphere electron-ion
potential with the hard-sphere radius equal to the atom radius of the excluded volume concept $r_a$ \cite{EbFortov}. It does not depend on the temperature of the system. For a typical atom radius of
$r_a=1.0\,a_\mathrm{B}$ 
we find $b_{2}^{\rm class}=b_{ea}^{\rm ex}=-3.1\times 10^{-25} \rm cm^3$.

\begin{figure}[htp]
 \includegraphics[width=0.37\textwidth,angle=-90]{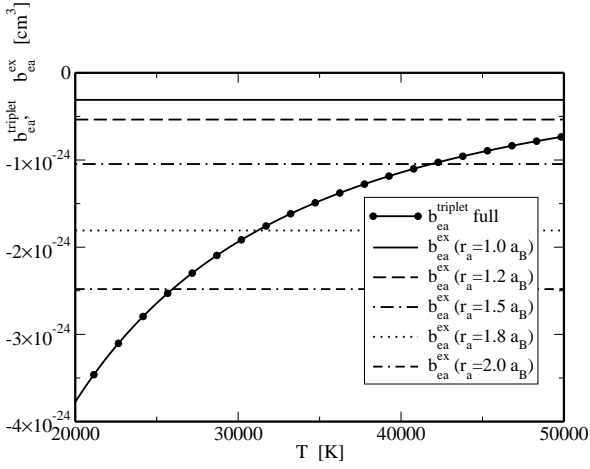}
\caption{Second virial coefficient in the triplet channel as function of temperature. The Beth-Uhlenbeck result (solid line with
circles) is compared to the excluded volume model for different values of $r_a$\label{fig:ex}}
\end{figure}

In Fig.~\ref{fig:ex} we show the second virial coefficient for the triplet state,
calculated by the Beth-Uhlenbeck formula using the experimental phase-shifts from \cite{schwartz} in
comparison with the excluded volume virial coefficient for different values of $r_a$. In the triplet state 
we have a strong repulsion between electrons and atoms, hence the hard-sphere potential can be expected to
give reasonable results. Because of the bound state formation, the singlet state can not be treated within the excluded volume approach.  
Note that in contrast to the excluded volume concept and the hard-sphere model,
our results indeed depend on the temperature. At high temperatures ($T\gtrsim 50000\,\text{K}$),
the second virial coefficient from our Beth-Uhlenbeck
calculation approaches the excluded volume virial coefficient for the atom radius $r_a=1.2\,a_B$. In this sense, the
Beth-Uhlenbeck using experimentally validated scattering phase-shifts provides a benchmark to the semi-empirical
excluded volume model.

Although the excluded volume concept is widely applied to take into account the
presence of atoms in the plasma, this method gives only approximate results. The dependence of the
atom subsystem on the plasma parameters was included in the confined atom model \cite{kraeft, rogers} 
due to an atomic radius.
These methods cannot cover numerous effects in the electron-atom interaction, such as the spin dependence,
scattering phase-shifts, and bound states which are included in the Beth-Uhlenbeck formula.    

% \begin{figure}[htp]
%  \includegraphics[width=0.37\textwidth,angle=-90]{relationofb2tobex.ps}
% \caption{Dependence of $b_{ea}^{\rm triplet,s-wave}/b_{ea}^{\rm ex}$ on the temperature for different radius of atom.\label{fig:ex1}}
% \end{figure}

\section{Generalized Beth-Uhlenbeck approach\label{sec:PauliBlock}}

The virial expansion can be extended to higher densities if the effects of the medium are taken into account. 
In particular, we outline the consequence of Pauli blocking on the two-particle
properties, that is of importance when the electrons become degenerate. There are other
medium effects such as screening, where the effective interaction potential between the electron and the atom is replaced by a screened
potential. The influence on the scattering processes using screened versions of the
Buckingham and the RDO models has been treated in Refs.~\cite{redmer1997,RDO} and will
not repeated here. A systematic approach to screening effects is given within the Green's function theory \cite{kraeft}.

We consider the two-particle effective  wave equation 
%As an example, we consider the two-particle effective wave equation
%\begin{widetext}
\begin{eqnarray}
\Bigg[\frac{p_1^2}{2m_1}+\Delta^{\rm SE}(p_1)+\frac{p_2^2}{2m_2}
+\Delta^{\rm SE}(p_2)\Bigg]\psi(p_1,p_2)+
\nonumber\\
\Big[ 1\pm f(p_1)\pm f(p_2)\Big]
\sum\limits_{p_1',p_2'}{V(p_1p_2,p_1'p_2')\psi(p_1',p_2')}
\nonumber\\
=E(P,T,\mu_c)\psi(p_1,p_2)\,,
\label{eq:l19}
\end{eqnarray}
%\end{widetext}
where $\Delta^{\rm SE}(p)$ denotes the self-energy shift, and $f(p)=1/\left(\exp{[\beta(\frac{p^2}{2m}-\mu)]\mp 1}\right)$ 
are the Bose and Fermi distributions. 
This approach has been applied to charged-particle systems
as detailed in Ref. \cite{kraeft} for the electron-ion system as well as for the electron-hole system. We
will use a similar approach for the $e-a$ problem.

The inclusion of self-energy, screening and Pauli blocking effects in the solution of the in-medium Schr\"odinger equation 
for the electron-ion system leads to non-ideality contributions. In particular, the Mott
effect is obtained, i.e. the dissolution of bound states in the continuum of scattering
states at increased densities.
The contribution of the energy shift of atomic levels on the thermodynamics of the dense
hydrogen was considered in Refs~\cite{ebeling,kraeft}.
The Pauli shift $\Delta^{\rm Pauli}=32 \pi n_e$ (in Rydberg units) at low temperatures
and at low densities and the Fock shift $\Delta^{\rm Fock}=-20 \pi n_e$, 
lead to modified behavior at high pressures. 
In Ref.~\cite{redmer}, the effects of 
Pauli blocking on transport properties of dense
plasma were investigated by  solving the 
thermodynamical T-matrix for
the electron-ion scattering for a 
separable electron-ion potential.

A generalized Beth-Uhlenbeck formula has been successfully elaborated for nuclear matter \cite{roepke_bu}. 
In particular, the Mott effect can be included so that the applicability of this approach is extended to the region 
where a quasiparticle description is possible, e.g. in the region of strong degeneracy. Analytical expressions are
derived for a separable potential approach where the in-medium T-matrix including Pauli blocking effects can be calculated.

We study the shift of the binding energy of H$^{-}$ as well as the modification of $e-a$ scattering phase-shifts
due to Pauli blocking. Our starting point is the effective Schr\"{o}dinger equation
for the $e-a$ problem
\begin{equation}
\big[\frac{p^2}{2m}+\Delta^{\rm SE}(p)\big]\psi(p)+\Big[1-f(p)\Big]\sum\limits_{p'}{V(p,p')\psi(p')}=E_{0}\psi(p) .
\label{eq:19}
\end{equation}
Medium effects are the self-energy (Fock term) $\Delta^{\rm SE}(p)$ and the Pauli
blocking term $[1-f(p)]$, that describes the occupation of phase space.

To investigate the Mott effect with respect to the formation of H$^{-}$, we investigate
the binding energy of the $e-a$ system as a function of density, i.e. 
the difference between the bound state energy and the continuum edge of scattering states. The self-energy
of electrons $\Delta^{\rm SE}(p)$ due to the electron-atom interaction shifts both 
the bound state energy as well as the scattering states, the net effect on the
ionization energy hence being zero.
The leading term is the  Pauli blocking term, that will be evaluated in the following. 

We determine the occupation number $f(p)$ in Eq.(\ref{eq:19})
via the chemical potential $\mu_e$ according to 
% 
% 
% 
%   are included 
% as shifts of the ionization potential, see Eq.~(\ref{eq:19}). 
% The self-energy effects can be absorbed
% in the effective electron mass. For comparison, the contribution of the Pauli blocking and the Fock term to the
% ground state shift of the hydrogen atom in the medium was considered in Ref.~\cite{ebeling}.
% The Pauli blocking shift for the limiting case at low temperature and low density  was found as
% $\Delta^{\rm Pauli}=32 \pi n_e$ in Rydberg units. The Fock shift was found of the same order
% as the Pauli blocking $\Delta^{\rm Fock}=-20 \pi n_e$. 
% 
% 
% However, the Fock shift plays no import
% role in the disappearance of the binding energy, because the shift of the bound state 
% is compensated by the lowering of the continuum edge, i.e. the Fock shift of
% the free electrons. Consequently, we focus only on the influence of the Pauli effect on the binding energy
% shift, which leads to the Mott effect in degenerate system.    
% 
% 
% Furthermore, we neglect the screening effect, 
% which can be studied by replacing the effective potential $V(p,p')$ by a screened potential. 
% The influence of the screening effect on the scattering processes using screened versions
% of the Buckingham and the RDO models has been treated in Ref.~\cite{RDOR}. Significant screening is obtained
% only at increased densities, but it is usually shadowed by degeneracy effects, such as Pauli
% blocking.
% 
% The chemical potential $\mu_e$ is given by the density according to
\begin{equation}
\int\limits_{0}^{\infty}\frac{d^3p}{(2\pi
\hbar)^3}\frac{1}{\exp{[\beta(\frac{p^2}{2m_e}-\mu_e)]+1}}=\frac{n_{e}}{2}\,.
\label{eq:20}
\end{equation}
\begin{figure}[htp]
  \begin{center}
\includegraphics[width=.37\textwidth,angle=-90]{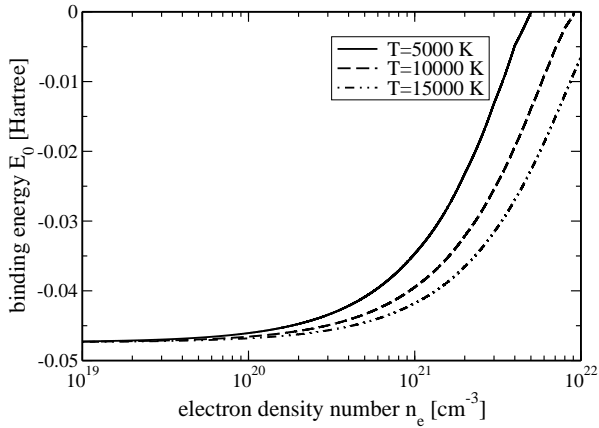}
\caption {H$^-$ binding energy in dependence of total electron number density $n_e$ for different temperatures $T$.
\label{fig:9}}
\end{center}
\end{figure}
\begin{figure}[htp]
\center
\includegraphics[width=.37\textwidth,angle=-90]{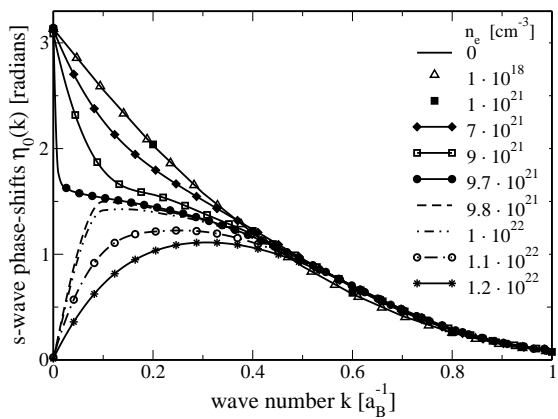}
\caption{$s$-wave scattering phase-shifts as function of wavenumber $k$ for different electron densities $n_e$ 
 at $T=10000$ K. For comparison, the low-density limit ($n_e=0$) is also shown.
\label{fig:10}}
\end{figure}

Using the parameters of the rank-one separable potential given in Tab. I, the
binding energies of the electron in the negative hydrogen ion have
been calculated in dependence of the total electron density and the temperature.
The numerical results for the in-medium binding energies are given in
Fig.~\ref{fig:9}. We see
that the binding energy is decreased with increasing electron density.
For $T=10000$ K, at the density exceeding the Mott density $n^{\rm Mott}_{e}=9.8\cdot10^{21}$ cm$^{-3}$, bound states cannot be formed.

The influence of the medium on the scattering phase-shifts is obtained by solving the
T-matrix including the Pauli blocking term. Results for different densities are presented
in Fig.~\ref{fig:10}. At the Mott density we
observe a jump of the in-medium $s$-wave phase-shift $\eta_0$
by $\pi$  
according to the Levinson theorem.

Using the density dependent phase-shifts and binding energies calculated from the in-medium Schr\"odinger
equation (\ref{eq:19}), 
we can calculate the scattering and bound parts of the second virial coefficient. 

The results are summarized in
Tabs. \ref{tab4} and \ref{tab5}. 
With increasing density the bound part of the second virial coefficient is decreasing because the binding energy becomes smaller due to the Pauli blocking and screening effects. 

It should be mentioned that the account of the self-energy $\Delta^{\rm SE}(p)$ would contribute to the chemical potential as determined 
by the normalization condition (\ref{eq:20}). If the density of the electrons is replaced by the density of quasiparticles that account for the self-energy
shift, we also have to modify the Beth-Uhlenbeck expression for the second virial
coefficient as shown in Ref.~\cite{roepke_bu}. The Beth-Uhlenbeck expression for the
second virial coefficient (\ref{eq:7}) used here is consistent with the single particle
contribution given by free particles as done here. In future investigations,
an improved treatment of density effect can be performed on the basis of 
quasiparticles and their corresponding screened interactions. 

\begin{table*}[htp]
\center
\caption{\label{tab4} Results only for $s$ -wave:the singlet scattering part of the second virial coefficient $\tilde{b}_{ae}^{\rm sc,singlet}(T,n_{e})$ for $e-H$ interaction 
for different electron number densities and temperatures. The phase-shifts are calculated with the separable model.}
\item[]
\begin{ruledtabular}
\begin{tabular}{c|c|c|c|c}
$T$, K & $n_{e}=10^{18} \rm cm^{-3}$& $n_{e}=10^{19} \rm cm^{-3}$&$n_{e}=10^{21} \rm cm^{-3}$& $n_{e}=5*10^{21} \rm cm^{-3}$\\
\hline 
5000 &-0.0505&-0.0504&-0.0464&-0.1012\\
 6000 &-0.0547&-0.0546&-0.0521&-0.0764\\
 7000 &-0.0584&-0.0583&-0.0571&-0.0750\\
 8000 &-0.0617&-0.0616&-0.0614&-0.0770\\
 9000 &-0.0647&-0.0646&-0.0653&-0.0797\\
 10000 &-0.0674&-0.0674&-0.0688&-0.0825\\
 11000 &-0.0700&-0.0700&-0.0708&-0.0851\\
 12000 &-0.0725&-0.0725&-0.0729&-0.0871\\
 13000 &-0.0748&-0.0748&-0.0755&-0.0894\\
 14000 &-0.0770&-0.0770&-0.0772&-0.0917\\
 15000 &-0.0792&-0.0792&-0.0794&-0.0939\\
\end{tabular}
\end{ruledtabular}
\end{table*}

\begin{table*}
\center
\caption{\label{tab5}The bound part of the second virial coefficient $\tilde{b}_{ae}^{\rm bound, singlet}(T,n_{e})$ for $e-H$ interaction for different electron number densities.} 

\begin{ruledtabular}
\item[]
\begin{tabular}{c|c|c|c|c}
$T$, K & $n_{e}=10^{17} \rm cm^{-3}$&$n_{e}=10^{19} \rm cm^{-3}$&$n_{e}=10^{21} \rm cm^{-3}$&$n_{e}=5*10^{21} \rm cm^{-3}$\\
\hline
5000&   4.8698&   4.8298&   2.1955&   0.2505\\

6000&   2.9688&   2.9506&   1.6384&   0.2865\\

7000&   2.0848&   2.0749&   1.3154&   0.3243\\

8000&   1.5993&   1.5932&   1.1080&   0.3529\\

9000&   1.3013&   1.2973&   0.9650&   0.3730\\

10000&   1.1034&   1.1006&   0.8610&   0.3864\\

11000&   0.9640&   0.9620&   0.7824&   0.3948\\

12000&   0.8615&   0.8599&   0.7210&   0.3996\\

13000&   0.7833&   0.7820&   0.6719&   0.4019\\

14000&   0.7219&   0.7209&   0.6318&   0.4023\\

15000&   0.6726&   0.6718&   0.5984&   0.4015\\
\end{tabular}
\end{ruledtabular}
\end{table*}

\section{Conclusions\label{sec:conclusion}}
For partially ionized plasmas, the cluster virial expansion for thermodynamic functions is considered. 
We focus on the contribution due to the  electron-atom interaction. 
With the help of the Beth-Uhlenbeck formula, the second virial coefficient in the electron-atom channel is related to phase-shifts and possible bound states in that channel. 
In contrast to former approaches, we give values for the second virial coefficient in the $e-a$ channel that are not based on any pseudopotential models 
but are directly related to measured data. 
Depending on the accuracy of  presently available experimental data, these
 exact results for the second virial coefficient can serve as a benchmark to test other more empirical approaches 
using pseudopotentials or related concepts to evaluate the thermodynamic properties of partially ionized plasmas.

From  the theoretical point of view, the $e-a$ interaction amounts to a three-particle problem.
At present, the most reliable numerical solutions are obtained from  variational calculations. After comparing these results with experimental scattering data, 
the second virial coefficient is presented in the range from 5 $\times 10^3$ to $10^5$ K.

The accurate calculation of the free energy excess due to electron-atom interaction is compared with excluded volume results 
that are widely used in the chemical model. This semi-empirical treatment contains the hard-core radius of the atom as an empirical parameter. 
Comparing the corresponding virial expansions, it is shown that a single parameter choice for the hard-core radius cannot reproduce the 
non-ideal contribution to the thermodynamic functions in a wide region of temperature.

We also considered different empirical pseudopotentials that can approximate these microscopic input quantities. 
In particular, a rank-two separable potential is introduced that fits the microscopic data. 
The advantage of a properly chosen pseudopotential is that higher order non-ideality terms with respect to the density 
can be calculated. For this, the on-shell properties in the two-body channel 
are no longer sufficient.

Going beyond the second virial coefficient, density effects such as self-energy shifts and Pauli blocking
have to be considered. In particular, we include Pauli blocking using the separable potential. 
In this way, we performe calculations for the density dependent second virial coefficient to cover electron densities up to near
solid densities $n_\mathrm{e}\simeq 10^{22}\,\text{cm}^{-3}$.

Our study of the contribution of the electron-atom interaction is a step to the systematic evaluation of the thermodynamics of 
partially ionized dense plasma where artificial parameters such as a hard-core radius are obsolete.
The main ingredient, the systematic transition from the physical picture to a chemical one, can be obtained from a quantum statistical approach.
The use of the technique of Green's functions allows for the account of higher order many-particle effects.

The generalization of the cluster-virial expansion, including more bound states as well as excited states, 
is straightforward. 
\section*{Acknowledgement}

This work is supported by the Deutsche Forschungsgesellschaft (DFG) under grant SFB 652 ``Strong Correlations and
Collective Phenomena in Radiation Fields: Coulomb Systems, Clusters, and Particles'' and the associated
Graduiertenkolleg. The work of Y.O. was partially supported by the scholarship of Deutscher Akademischer Austausch Dienst (DAAD). 
C.F. acknowledges support by the Alexander von Humboldt-Foundation and hospitality by Lawrence Livermore National
Laboratory.

\bibliography{literature}
\begin{appendix}
    \section{Cluster virial expansion and thermodynamic functions \label{app:fugacExp}}
  We consider a system of interacting electrons ($e,s_e=1/2$), ions (protons $i, s_i=1/2$), and atoms ($a$) in the singlet state ($s_{a,\rm singlet}=0$) and in the triplet state ($s_{a,\rm triplet}=1$). We neglect the hyperfine splitting.

 Within the cluster virial expansion, the grand canonical partition function,
  $\Omega(z_e,z_i,z_a,T,V)$ is a function of
  the fugacities, $z_c=e^{\beta (\mu_c-E_c^{(0)})}$ (see Sec. II),
the temperature $T$, and the system's volume $V$. We expand up to the second order in the fugacities:
  \begin{eqnarray}
    \Omega(z_c,T,V)&=&1+\sum_{c=e,i,a}z_c\Omega_{c}(T,V)
\nonumber\\
&+&\sum_{c,d}z_cz_d\Omega_{cd}(T,V)+\mathcal{O}(z_c^3)~.
    \label{eqn:app:GKP_cvirialexp}
  \end{eqnarray}
  Here, we have introduced the single-particle partition functions $\Omega_c(T,V)=(2s_c+1)V/\Lambda_c^3$
  and the two-particle partition functions $\Omega_{cd}(T,V)$ that are related to the
  interaction, $\Lambda_c=\left( 2\pi\hbar^2/m_c k_\mathrm{B}T
  \right)^{1/2}$ is the thermal wavelength of species $c$.
  The two-particle partition will be related below to the second virial coefficient.
  From the partition function, we can directly derive the pressure $P$ in the system,
  \begin{equation}
    P(z_c,T,V)=\frac{k_\mathrm{B}T}{V}\ln \Omega(z_c,T,V)~.
    \label{eqn:app:Pressure}
  \end{equation}
  Replacing the partition function by Eq.~(\ref{eqn:app:GKP_cvirialexp}) and expanding again up to second order in $z_c^2$, we arrive at
  \begin{eqnarray}
    \frac{P(z_c,T,V)}{k_\mathrm{B}T}&=&\sum_{c}\frac{2s_c+1}{\Lambda_c^3}z_c
\nonumber\\
&+&\sum_{cd}z_cz_d\frac{1}{V}(\Omega_{cd}-\frac{1}{2}\Omega_c\Omega_d)+
    \mathcal{O}(z_c^3)~
\nonumber\\
&=&\sum_{c}\frac{2s_c+1}{\Lambda_c^3}(z_c+\sum_{d}z_cz_d\tilde{b}_{cd})+\mathcal{O}(z_c^3)~.
\nonumber\\    
\label{eqn:app:Pressure_zQ}
  \end{eqnarray}

 We have introduced the second virial coefficients $\tilde{b}_{cd}=\frac{\Lambda_c^3}{2s_c+1}(\Omega_{cd}-\frac{1}{2}\Omega_c\Omega_d)/V$
  that are related to the symmetric expressions $b_{cd}=\frac{\Lambda_d^3}{2s_d+1}\tilde{b}_{cd}$.

The thermodynamic functions of the partially ionized hydrogen plasma are derived from the grand
  canonical potential Eq.~(\ref{eqn:app:GKP_cvirialexp}). First, we evaluate the number densities of each
  component,
  \begin{eqnarray}
    n_c=z_c\left( \frac{\partial}{\partial z_c} \frac{\ln \Omega(z_c,T,V)}{V} \right)_{T,V}
\nonumber\\
= z_c\left( \frac{\partial}{\partial z_c} \frac{P(z_c,T,V)}{k_\mathrm{B}T} \right)_{T,V}~.
    \label{eqn:app:partial_densities}
  \end{eqnarray}
  In terms of the second virial coefficient, evaluation of the derivatives with respect to the fugacity yields
\begin{equation}
    \label{eqn:app:ne}
    n_c = \frac{2s_c+1}{\Lambda_c^3}\left(z_c + 2 \sum_{d}z_cz_d\frac{2s_d+1}{\Lambda_d^3}b_{cd}+\mathcal{O}(z_c^3)\right) 
\end{equation}

  In the first terms of equations (\ref{eqn:app:ne}),
  we recognize the partial densities of the ideal (non-interacting) system, $n_{c,\mathrm{id}}=\left( 2s_c+1\right) z_c/\Lambda_c^3$.

  Next, we evaluate the entropy density of the partially ionized plasma ($z_c$ is a function of $\mu_c$ and $T$), 
  \begin{eqnarray}
    &&\frac{S(z_c,T,V)}{V}= \left( \frac{\partial P(z_c,T,V)}{\partial T} \right)_{z_c,V} 
\nonumber\\
&=& \frac{\partial}{\partial T} \left( k_\mathrm{B}T\frac{\ln \Omega(z_c,T,V)}{V} \right)_{z_c,V} 
\nonumber\\
&=&\frac{P(z_c,T,V)}{T}
+\frac{k_\mathrm{B}T}{V}\left( \frac{\partial \ln\Omega(z_c,T,V)}{\partial T} \right)~.
    \label{eqn:app:entropydensity_010}
  \end{eqnarray}
  After some lengthy manipulations we obtain the result
  \begin{equation}
    \frac{5}{2}\frac{P}{T} - k_\mathrm{B}\sum_{c}n_c\ln z_c
    + k_\mathrm{B}T\sum_{cd}\frac{(2s_c+1) z_cz_d}{\Lambda_c^3}\frac{\partial \tilde{b}_{cd}(T)}{\partial T} 
    \label{eqn:app:entropydensity_020}
  \end{equation}

  The density of internal energy follows from the relation
  \begin{eqnarray}
    \label{eqn:app:internalenergy_relation_s_n_P}
 \frac{U(z_c,T,V)}{V}=\frac{TS}{V} + \sum_{c} \mu_cn_c - P&&
   \nonumber\\
 =\frac{3}{2}P - \sum_{c}n_cE_c
+k_BT^2\sum_{cd}\frac{(2s_c+1) z_cz_d}{\Lambda_c^3}\frac{\partial \tilde{b}_{cd}}{\partial T}&&
  \end{eqnarray}
  And finally, we find for the free energy density $f(n_c,T)=F/V$ after eliminating $z_c$ with Eq.(\ref{eqn:app:ne})
  \begin{eqnarray}
   f(n_c,T)&=&\frac{U}{V}-\frac{TS}{V}
\nonumber\\    
&=&\sum_{c}\mu_cn_c- P
\nonumber\\    
&=&k_BT\sum_{c}n_c[\ln\frac{\Lambda_c^3n_c}{2s_c+1}-1+\beta E_c^{(0)}]
\nonumber\\
&-&k_BT\sum_{cd}n_cn_db_{cd}
    \label{eqn:app:freeenergydensity}
    \end{eqnarray}
that coincides with the expression (\ref{eq:3}).

% \textbf{The different form of the density virial expansion for the equation of state in literature \cite{landau,BU,huang} can mislead, but the form of the expansion depends on the definition of the second virial coefficient. The second virial coefficient of the density expansion $B(T)$  in the book by Landau and et al. \cite{landau} is related to $b_{cd}$ from present paper as $B(T) =-b_{cd}$, the second virial cofficient of the fugacity expansion $b$ in the Ref.\cite{horowitz} is related to $\tilde{b_{cd}}$ from present paper as $b=\tilde{b_{cd}}$ } 

\section{T-matrix for separable potential\label{app:Tmatrix}}

We consider the low density limit $n_\mathrm{e}\Lambda_\mathrm{e}^3\ll 1$, where the T-matrix equation \cite{Schwinger} for a separable potential reads:
\begin{eqnarray}
 &&T(p,p',E)=V(p,p')
\nonumber\\
&+&\sum_{p^{''}}V(p,p^{``})\frac{1}{\frac{p^{''2}}{2m_r}-E}T(p^{''},p^{'},E).
\label{eq:s4}
\end{eqnarray}
If we consider a rank-two separable potential (\ref{eq:s1}), we obtain the following expression for the T-matrix: 
\begin{widetext}
\begin{eqnarray}
 T(p,p^{'},E)&=&\lambda_1w_1(p)w_1(p')+\lambda_2w_2(p)w_2(p')+
\lambda_1w_1(p)\int\frac{d^3p''}{(2\pi)^3}w_1(p'')\frac{1}{\frac{p^{''2}}{2m_r}-E}T(p'',p',E)
\nonumber\\
&&+\lambda_2w_2(p)\int\frac{d^3p''}{(2\pi)^3}w_2(p'')\frac{1}{\frac{p^{''2}}{2m_r}-E}T(p'',p',E)
\nonumber\\
&&=c_{11}w_1(p)w_1(p')+c_{22}w_2(p)w_2(p')+c_{12}w_1(p)w_2(p')+c_{21}w_2(p)w_1(p'), 
\label{eq:s5}
\end{eqnarray}
\end{widetext}
with 
\begin{eqnarray}
c_{ij}&=&\lambda_i\delta_{ij}
+\lambda_i\int\frac{d^3p''}{(2\pi)^3}w_i(p'')\frac{1}{\frac{p^{''2}}{2m_r}-E}(c_{1j}w_1(p'')
\nonumber\\
&+&c_{2j}w_2(p'')) 
\end{eqnarray}

This set of equations can be simplified if we introduce the integrals $I_{ij}(E)=\int\frac{d^3p''}{(2\pi)^3}\frac{1}{\frac{p^{''2}}{2m_r}-E}c_{ij}w_i(p'')w_j(p'')$:  
\begin{eqnarray}
 c_{ij}=\lambda_i\delta_{ij}+\lambda_iI_{i1}(E)c_{1j}+\lambda_iI_{i2}(E)c_{2j}
\end{eqnarray}

Finally, after some mathematics, we obtain the T-matrix in the following form: 
\begin{eqnarray}
T(p,p',E)&=&\frac{1}{\rm Det(E)}\Big\{(\lambda_1\big[1-\lambda_2I_{22}(E)\big]w_1(p)w_1(p')
\nonumber\\
&+&\lambda_2\big[1-\lambda_1I_{11}(E)\big]w_2(p)w_2(p')
\nonumber\\
&+&\lambda_1\lambda_2I_{12}(E)w_1(p)w_2(p')
\nonumber\\
&+&\lambda_1\lambda_2I_{21}(E)w_2(p)w_1(p')\Big\}
\end{eqnarray}
where
\begin{eqnarray}
 {\rm Det(E)}&=&\Big[1-\lambda_1I_{11}(E)\Big]\Big[1-\lambda_2I_{22}(E)\Big]
\nonumber\\
&-&\lambda_1\lambda_2I_{12}(E)I_{21}(E).
\end{eqnarray}
 
Using properties of the T-matrix, the binding energy $E_0$ can be obtained from the equation ${\rm Det(E_0)}=0$.  
\end{appendix}

\end{document}